\documentclass[aps,prl,twocolumn,showpacs,superscriptaddress]{revtex4}

\usepackage{graphicx}  
\usepackage{dcolumn}   
\usepackage{bm}        
\usepackage{amssymb}   
\usepackage{amsmath}
\usepackage{units}
\usepackage{times}
\usepackage[ansinew]{inputenc}

\begin{document}

\title{The energy resolution function of a tunnel junction}

\author{Christian R. Ast}
\email[Corresponding author; electronic address:\ ]{c.ast@fkf.mpg.de}
\affiliation{Max-Planck-Institut f\"ur Festk\"orperforschung,
70569 Stuttgart, Germany}
\author{Berthold Jäck}
\affiliation{Max-Planck-Institut f\"ur Festk\"orperforschung,
70569 Stuttgart, Germany}
\author{Jacob Senkpiel}
\affiliation{Max-Planck-Institut f\"ur Festk\"orperforschung,
70569 Stuttgart, Germany}
\author{Matthias Eltschka}
\affiliation{Max-Planck-Institut f\"ur Festk\"orperforschung,
70569 Stuttgart, Germany}
\author{Markus Etzkorn}
\affiliation{Max-Planck-Institut f\"ur Festk\"orperforschung,
70569 Stuttgart, Germany}
\author{Joachim Ankerhold}
\affiliation{Institut für Komplexe Quantensysteme and IQST, Universität Ulm, 89069 Ulm, Germany}
\author{Klaus Kern}
\affiliation{Max-Planck-Institut f\"ur Festk\"orperforschung,
70569 Stuttgart, Germany}
\affiliation{Institut de Physique de la Matière Condensée, Ecole Polytechnique Fédérale de Lausanne, 1015 Lausanne, Switzerland}

\date{\today}

\begin{abstract}
The tunnel junction between tip and sample in a scanning tunneling microscope is an ideal platform to access the local density of states in the sample through the differential conductance. We show that the energy resolution that can be obtained is principally limited by the electromagnetic interaction of the tunneling electrons with the surrounding environmental impedance as well as the capacitative noise of the junction. The parameter tuning the sensitivity to the environmental impedance is the capacitance of the tunnel junction. The higher the junction capacitance, the less sensitive the tunnel junction to the environment resulting in better energy resolution. Modeling this effect within $P(E)$-theory, the $P(E)$-function describes the probability for a tunneling electron to exchange energy with the environment and can be regarded as the resolution function of the tunnel junction. We experimentally demonstrate this effect in a scanning tunneling microscope with a superconducting aluminum tip and a superconducting aluminum sample at a base temperature of 15\,mK, where it is most pronounced.
\end{abstract}


\maketitle
Scanning tunneling spectroscopy has evolved into one of the most versatile tools to study the electronic structure in real space with atomic precision \cite{wiesendanger_spin_2009,bode_spin-polarized_2003,leemput_scanning_1992}. The differential conductance measured through the tunneling contact directly accesses the local density of states of the sample. With the growing interest in phenomena with extremely sharp spectral features on smaller and smaller energy scales, the demand for higher and higher spectroscopic energy resolution increases. Examples are the Kondo effect \cite{zhang_temperature_2013}, Yu-Shiba-Rusinov states \cite{yazdani_probing_1997,franke_competition_2011}, Majorana fermions \cite{nadj-perge_observation_2014}, or the Josephson effect \cite{naaman_fluctuation_2001,kimura_josephson_2009,jack_nanoscale_2015}, just to name a few. Aside from the obvious strategy of lowering the temperature to increase the energy resolution \cite{tersoff_theory_1985}, superconducting tips have successfully been employed to circumvent the broadening effects of the Fermi function in the tunneling process and greatly improve the energy resolution \cite{pan_vacuum_1998,franke_competition_2011}. However, at low temperatures, other energy scales such as the charging energy $E_C$ of the tunnel junction may become important limiting the maximum achievable energy resolution. The question arises whether the tunneling process encompasses an intrinsic resolution limit, however small it may be, which cannot be overcome.

\begin{figure}
\centerline{ \includegraphics[width = \columnwidth]{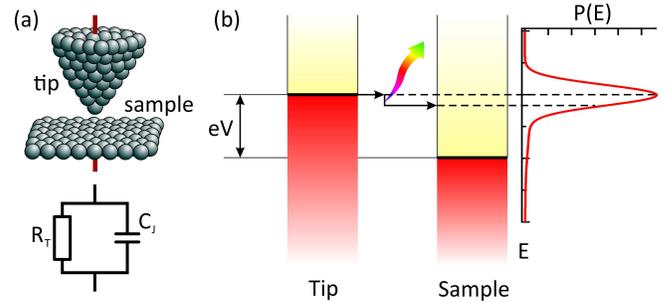}}
\caption{(a) Schematic drawing of an STM tunnel junction consisting of tip and sample. The equivalent circuit diagram is represented by a tunneling resistor $R_T$ and the junction capacitance $C_J$. (b) Schematic energy diagram showing the energy loss of an electron tunneling in an STM from the tip to the sample. Interacting with the surrounding environment, the electron loses energy according to the probability given by the $P(E)$-function.}   \label{fig:schema}
\end{figure}

The principal properties of a tunnel junction can in many cases be satisfactorily described by the tunneling resistance \cite{chen_introduction_2008}. However, there are instances, where the capacitance of a tunnel junction has to be taken into account as shown in Fig.\ \ref{fig:schema}(a). This is the case if the capacitance is small and the fact that the transport of charges across the junction is actually quantized becomes important \cite{delsing_effect_1989,devoret_effect_1990,averin_incoherent_1990}. An electron tunneling from one capacitor plate to the other leads to a ``sudden'' change of the electric field in the close vicinity of the tunnel junction. A back action of the surrounding electromagnetic impedance may lead to the emission or absorption of a photon by the tunneling electron accompanied by a loss or gain in energy, respectively. This is schematically shown in Fig.\ \ref{fig:schema}(b) for an STM, where the electron tunnels from the tip to the sample and loses a photon to the environment. In addition, the thermal noise of the junction capacitance becomes appreciable for small capacitance values. These effects may be small, but they are non-negligible at low temperatures, especially in the context of ultimate energy resolution.

The so-called $P(E)$-theory quantifies the energy exchange with the environment, where the $P(E)$-function describes the probability of a tunneling electron to emit or absorb a photon to or from the environment \cite{devoret_effect_1990,averin_incoherent_1990,ingold_cooper-pair_1994,ingold_charge_1992,ingold_finite-temperature_1991}. The $P(E)$-theory has already been successfully applied in many instances, where tunneling electrons interact with the surrounding electromagnetic impedance. Examples are the Josephson effect in the charge tunneling regime \cite{ingold_cooper-pair_1994,hofheinz_bright_2011}, as well as general dynamic Coulomb blockade effects \cite{delsing_effect_1989,devoret_effect_1990,brun_dynamical_2012,serrier-garcia_scanning_2013}. These two examples represent special conditions: tunneling of Cooper pairs and tunneling in a high-impedance environment, respectively. Nevertheless, the $P(E)$-theory should also apply to a general tunnel junction. Pekola \textit{et al.} have very nicely described a contribution to the superconducting density of states (i.\ e.\ the Dynes parameter) by environmentally assisted tunneling through a normalconductor-insulator-superconductor junction \cite{pekola_environment-assisted_2010}. However, they did not generalize their findings to the general properties of a tunnel junction.

In this Letter, we show that the photon exchange of tunneling electrons with the surrounding environment in conjunction with the capacitative junction noise represents a principal limit of the energy resolution in spectroscopic measurements using tunnel junctions. In this regard, the $P(E)$-function represents the resolution function of a particular tunnel junction. Using a scanning tunneling microscope operating at 15\,mK \cite{assig_10_2013}, we independently characterize the $P(E)$-function of a superconductor-vacuum-superconductor tunnel junction through its direct relation with the Josephson effect. Subsequently, we demonstrate the impact of the $P(E)$-function on the tunneling process by measuring the superconducting quasiparticle density of states of the same tunnel junction. We find excellent quantitative agreement of our spectra from different tunnel junctions with the model calculations including $P(E)$-theory. This leads us to conclude that the $P(E)$-function plays a ubiquitous role as a resolution function in the tunneling process, in particular for energy scales at or below 1\,meV.

In order to consider the effects of the capacitance into the tunneling Hamiltonian, a small operator, which transfers a quantized amount of charge from one side to the other, has to be included \cite{devoret_effect_1990}. The resulting tunneling probability $\overrightarrow{\Gamma}(V)$ from tip to sample as a function of applied bias voltage is given by \cite{devoret_effect_1990,odintsov_effect_1988}:
\begin{widetext}
\begin{equation}
\overrightarrow{\Gamma}(V)=\frac{1}{e^2R_T}\int\limits^{\infty}_{-\infty}\int\limits^{\infty}_{-\infty}dEdE'n_\text{t}(E)n_\text{s}(E'+eV)
f(E)[1-f(E'+eV)]P(E-E')
\label{eq:tunprob}
\end{equation}
\end{widetext}
Here, $R_T$ ist the tunneling resistance, $f(E)=1/(1+\exp(E/k_BT))$ is the Fermi function, and $n_\text{t}$, $n_\text{s}$ are the densities of states of tip and sample, respectively. By exchanging electrons and holes in Eq.\ \ref{eq:tunprob}, the other tunneling direction $\overleftarrow{\Gamma}(V)$ from sample to tip can be obtained. Eq.\ \ref{eq:tunprob} differs from the standard expression of the tunneling probability by the convolution with the $P(E)$-function \cite{tersoff_theory_1985,bardeen_tunnelling_1961}. If we set $P(E)=\delta(E)$, which means that there is no energy exchange with the environment and no capacitative noise, the standard expression for the tunneling current is recovered \cite{chen_theory_1988}. It can be clearly seen that the convolution with the $P(E)$-function results in a broadening of the spectral features in the density of states. The current $I(V)$, which is measured through the tunnel junction as a function of applied bias voltage $V$, is the difference of the tunneling probabilities in the forward $\overrightarrow{\Gamma}(V)$ and the backward direction $\overleftarrow{\Gamma}(V)$. The tunneling current $I(V)$ is then:
\begin{equation}
I(V) = e\left(\overrightarrow{\Gamma}(V)-\overleftarrow{\Gamma}(V)\right)
\label{eq:iv}
\end{equation}
The effect of the capacitance in the tunnel junction and the interaction of the tunneling electrons with the surrounding environmental impedance has been modeled within the framework of $P(E)$-theory, where the $P(E)$-function describes the probability for a tunneling electron to exchange energy with the environment. It is commonly defined through the equilibrium phase correlation function $J(t)$ as the Fourier transform of $\exp[J(t)]$ \cite{ingold_cooper-pair_1994,ingold_charge_1992}. To account for the different dissipation channels, we define $J(t)=J_0(t)+J_N(t)$, where $J_0(t)$ describes the phase correlation of the environmental impedance and $J_N(t)$ captures the low frequency capacitative thermal noise in the junction. This allows us to calculate the $P_0(E)$-function for the environmental impedance and the $P_N(E)$-function for the capacitative noise separately. They can be combined to the total $P(E)$-function through a convolution \cite{si}.

Because the correlation function is difficult to calculate directly, we choose an implicit definition for $P_0(E)$ \cite{ingold_finite-temperature_1991}:
\begin{equation}
P_0(E)=I(E)+\int\limits^{\infty}_{-\infty}d\omega K(E,\omega)P_0(E-\hbar\omega)
\end{equation}
where the two functions $I(E)$ and $K(E,\omega)$ are defined in the supplementary information \cite{si}. The $P_0(E)$-function is parameterized by the temperature $T$, as well as the junction capacitance $C_J$, and the surrounding impedance $Z(\omega)$, which together form the total impedance:
\begin{equation}
Z_T(\omega)=\frac{1}{i\omega C_J + Z^{-1}(\omega)}.
\label{eq:zt}
\end{equation}
In our STM, the tip acts as a monopole antenna, whose impedance can be modeled in analogy to an infinite transmission line impedance \cite{jack_nanoscale_2015,ingold_cooper-pair_1994,si}. The fit parameters are the principal resonance frequency $\omega_0$ and a damping factor $\alpha$. The dc resistance $Z(0)$ is fixed at the vacuum impedance value of 376.73\,$\Omega$. The $P_N(E)$-function is modeled by a normalized Gaussian of width $\sigma=\sqrt{2E_Ck_BT}$ to account for the thermal voltage noise on the junction capacitor. Because both tip and sample are superconducting, we define the charging energy $E_C=Q^2/2C_J$ using the charge of a Cooper pair ($Q=2e$) \cite{ingold_cooper-pair_1994}, which makes it four times higher than the charging energy for a singly charged quasiparticle. Including the capacitative noise has proven essential in previous descriptions of the tunneling current as well \cite{jack_nanoscale_2015}.

The experiments were carried out in an STM operating at a base temperature of 15\,mK \cite{assig_10_2013}. We use an Al tip and an Al(100) sample \cite{si}, which is superconducting at 15\,mK (transition temperature $T_C=1.1\,$K). Aluminum has a very BCS-like density of states with minimal intrinsic broadening \cite{giaever_tunneling_1962}, which is why it is an excellent material for demonstrating the broadening effects in a tunnel junction, as we will show in the following.

In order to demonstrate the influence of the $P(E)$-function on density of states measurements, we have to independently determine the $P(E)$-function from a separate measurement. Since every tip in an STM is slightly different, we have to determine the $P(E)$-function for every tip separately. As both tip and sample are superconducting at 15\,mK, the most straightforward way to experimentally determine the $P(E)$-function is through the Josephson effect. In the sequential charge tunneling regime, where the charging energy $E_C$ is larger than the Josephson energy $E_J$ (which is commonly the case in a standard STM setup), the current-voltage characteristics is given by \cite{ingold_cooper-pair_1994}:
\begin{equation}
I(V)=\frac{\pi e E^2_J}{\hbar}\left[P(2eV)-P(-2eV)\right]
\label{eq:ivj}
\end{equation}
The Josephson energy $E_J$, can be regarded as a scaling parameter here \cite{ambegaokar_tunneling_1963}. In this sense, the $I(V)$ measurement of the Josephson effect is a direct measure of the $P(E)$-function \cite{ingold_charge_1992,ingold_cooper-pair_1994}.

\begin{figure}
\centerline{ \includegraphics[width = 1.0\columnwidth]{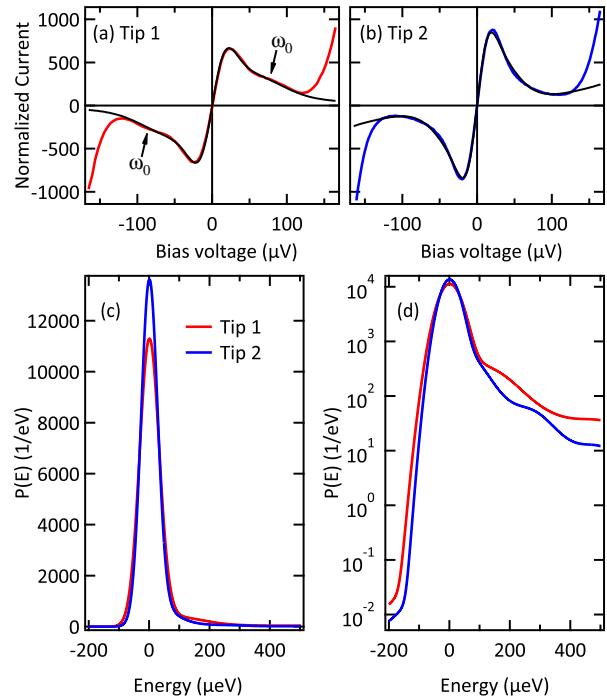}}
\caption{$I(V)$-characteristics of the Josephson effect for two different Al tips on an Al(100) sample. The tunnel junction in panels (a) and (b) have a capacitance of 3.5\,fF and 7\,fF, respectively, which means that the tunnel junction in panel (a) is more sensitive to the surrounding environment. Therefore, the principal impedance resonance $\nu_0$ is visible in panel (a) and not in panel (b). The fit using $P(E)$-theory (black lines) is in excellent agreement with the data. The $P(E)$-functions extracted from the fits in (a) and (b) are shown on a linear scale in (c) and on a logarithmic scale in (d). The asymmetry of the $P(E)$-function is clearly visible. } \label{fig:jos}
\end{figure}

We have measured the $I(V)$-characteristics of the Josephson effect for two different aluminum tips, which is shown in Fig.\ \ref{fig:jos}. The tunneling conditions were such that for both tips the current setpoint was 5\,nA at a voltage of 1\,meV and 2\,meV for tip 1 and 2, respectively. For better comparison, the current was divided by $\pi e E^2_J/\hbar$. The general features of the Josephson effect in the sequential charge tunneling regime are visible, however, the peak in Fig.\ \ref{fig:jos}(b) is somewhat higher and sharper than in panel (a). In addition, the spectrum in panel (a) shows a broad peak of the principal antenna resonance $\omega_0$. The fits using Eq.\ \ref{eq:ivj} are shown as black lines \cite{si}. They agree well with the measured data. The most significant difference between the two tunnel junctions is that the junction in the Fig.\ \ref{fig:jos}(a) has a capacitance of $C_J=3.5\pm0.2$\,fF, while in Fig.\ \ref{fig:jos}(b) the capacitance is $C_J=7.0\pm0.1$\,fF. The lower the junction capacitance value $C_J$ is, the more sensitive the junction will be to the environment $Z(\omega)$ (cf.\ Eq.\ \ref{eq:zt}).

The actual shape of the $P(E)$-function for the two tips is shown in Fig.\ \ref{fig:jos}(c) and on a semi-log scale in (d). The full width at half maximum (FWHM) for tip 1 and 2 is 77.2\,$\mu$eV and 65.4\,$\mu$eV, respectively, which will have a non-negligible effect on the superconducting density of states. In addition, the $P_0(E)$-function obeys the detailed balance symmetry \cite{ingold_finite-temperature_1991}, $P(E)=\exp(E/k_BT)P(-E)$, which makes the function inherently asymmetric as can be clearly seen in Fig.\ \ref{fig:jos}(d) even after the convolution with the capacitative noise. With the well-defined $P(E)$-function, we can look at its impact on the details of a quasi-particle spectrum.

\begin{figure}
\centerline{ \includegraphics[width = \columnwidth]{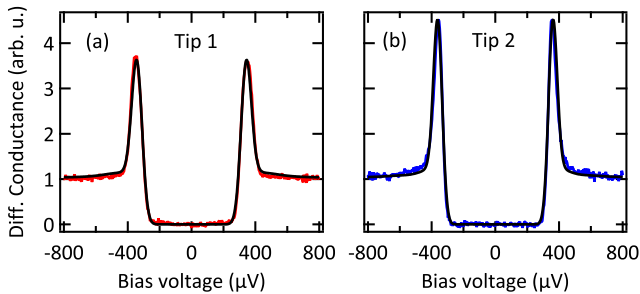}}
\caption{Differential conductance spectra of the superconducting densities of states for two different Al tips on an Al(100) sample. In order to suppress subgap features, we have measured at low transmission (stabilization at 2\,meV and 50\,pA). The fits are shown as black lines with excellent agreement. The superconducting density of states of both tip and sample was modeled by the simple BCS-model without any additional broadening parameters.} \label{fig:didv}
\end{figure}

The differential conductance $dI/dV$ spectra measured with the two tips as a function of bias voltage $V$ are shown in Fig.\ \ref{fig:didv} \cite{li}. Because both the tip and the sample are superconducting, the apparent gap in the spectrum has a width of $2(\Delta_\text{t}+\Delta_\text{s})$. The tip gap $\Delta_\text{t}$ can be slightly smaller than the bulk value of 180\,$\mu$eV. The most noticeable difference between the two spectra are the height of the coherence peaks. We fit the two spectra with the differential conductance model obtained from the derivative of Eq.\ \ref{eq:iv} in combination with Eq.\ \ref{eq:tunprob}. For the superconducting density of states in tip $n_{\text{t}}$ and sample $n_{\text{s}}$, we use the simple BCS-model \cite{bardeen_theory_1957}, explicitly neglecting any intrinsic broadening (e.\ g.\ Dynes parameter $\Gamma$ \cite{dynes_direct_1978}):
\begin{equation}
n_{\text{t,s}}(E)=n_0\Re\left[\frac{E}{\sqrt{E^2-\Delta_{\text{t,s}}^2}}\right]
\end{equation}
For the $P(E)$-function we use the same values that have been obtained from the fit to the data in Fig.\ \ref{fig:jos}, which means that the number of fit parameters is reduced to the value of the superconducting gap $\Delta$ and an overall scaling factor including the tunneling resistance $R_T$. We find excellent agreement for both spectra using the corresponding $P(E)$-function and with gap values of $\Delta_\text{t}=160\pm2\,\mu$eV for tip 1 and $\Delta_\text{t}=180\pm2\,\mu$eV for tip 2. The sample gap is set to the bulk value $\Delta_\text{s}=180\,\mu$eV. In both cases, the height and the shape of the coherence peaks are quantitatively well reproduced.

By contrast, disregarding the $P(E)$-function and fitting the spectra in Fig.\ \ref{fig:didv} with the Dynes equation \cite{dynes_direct_1978} to account for the broadening, does not give a satisfactory fit at all (see Supplementary Information \cite{si}). The height reduction in the coherence peaks has to be absorbed into the empirical broadening parameter $\Gamma$. This leads to the accumulation of quasiparticle spectral weight inside the gap, which is not observed experimentally. Therefore, we attribute the much better fit of the $P(E)$-function broadening to its inherent asymmetry strengthening the validity of our approach.

According to the $P(E)$-theory, the exchange of energy with the environment during the tunneling process should be an ubiquitous phenomenon \cite{devoret_effect_1990}. In the majority of cases, it has been discussed in the context of dynamical Coulomb blockade \cite{delsing_effect_1989,devoret_effect_1990,brun_dynamical_2012,serrier-garcia_scanning_2013} as well as the Josephson effect in the sequential charge tunneling regime \cite{ingold_cooper-pair_1994,averin_incoherent_1990,hofheinz_bright_2011}. While in these cases, the role of the $P(E)$-theory is obvious, in the present case as a resolution function in \textit{every} tunneling spectrum its role is more subtle, but non-negligible as we will show in the following. Neglecting the asymmetric shape for a moment and looking at the general broadening effect of the $P(E)$-function, we find that the FWHM of the $P(E)$-function is dominated by the capacitative noise and is essentially determined by the junction capacitance $C_J$ as well as the temperature $T$.

\begin{figure}
\centerline{\includegraphics[width = 0.8\columnwidth]{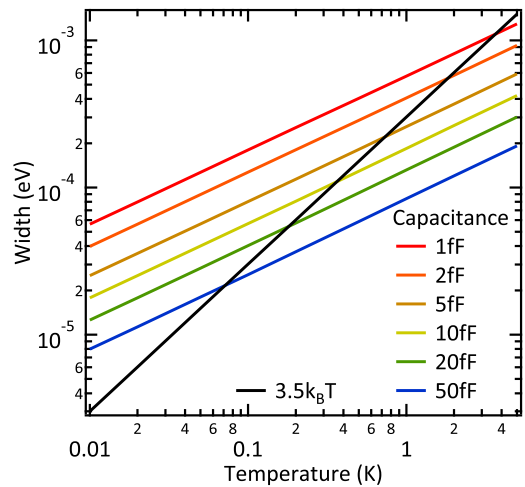}}
\caption{Energy broadening due to $P(E)$-broadening from the environment in comparison to thermal broadening from the Fermi functions as a function of temperature. The $P(E)$-broadening is shown for typical junction capacitances that can be found in an STM tunnel junction. For the total energy broadening both contributions have to be combined.} \label{fig:res}
\end{figure}

The FWHM of the $P(E)$-function for typical low temperatures (0.01 to 5\,K) and capacitances (1 to 50\,fF) is shown in Fig.\ \ref{fig:res}. For comparison, the thermal broadening $\Delta E_{\text{therm}}$ of differential conductance spectra due to the Fermi function is also shown as a black line ($\Delta E_{\text{therm}}=3.5k_BT$). While the thermal Fermi function broadening depends linearly on temperature, we find an overall empirical relation for the effective energy resolution due to $P(E)$-broadening, which is a function of temperature and capacitance $\Delta E_{\text{P(E)}}=\gamma\sqrt{2E_Ck_BT}$. The coefficient $\gamma$ has an average value of $\gamma=2.45\pm0.1$, keeping in mind that the capacitative noise is the dominant contribution to the FWHM. The $P_0(E)$-function changes the coefficient $\gamma$ slightly depending on the actual values of the parameters. This means that for low enough temperature, the $P(E)$-broadening will eventually be the dominant contribution to the resolution limit, regardless of whether the tip and/or sample are superconducting or not. We note that this empirical equation holds for capacitive noise from Cooper pairs. For noise from quasiparticles, we expect the $P(E)$-broadening to be reduced by about one half. At or below 1\,K, the $P(E)$-broadening definitely has to be taken into account when optimizing the energy resolution. The optimizing strategy will be to increase the junction capacitance by appropriate \textit{ex situ} tip shaping on a macroscopic scale (up to mm-scale). Increasing the junction capacitance will increase the crosstalk between tip and sample, so that a trade-off between energy resolution and STM performance will have to be made.

Due to the asymmetry of the $P(E)$-function, the spectral features in the density of states will not only be broadened, but may also change shape, which can have a strong influence on the interpretation of experimental data. The asymmetry evens out for higher temperatures, but at low temperatures, it has to be considered as can be seen in the fits of the differential conductance spectra in Fig.\ \ref{fig:didv}. If a symmetric broadening had been sufficient to fit these spectra, a Dynes fit would likely have sufficed. We expect the $P(E)$-broadening to be most significant on intrinsically sharp spectral features, such as coherence peaks of a superconducting gap. In addition, sharp Kondo peaks with a low Kondo temperature on the order of 1\,K may show an effectively higher Kondo temperature when the $P(E)$-broadening is not taken into account. Also, Yu-Shiba-Rusinov states, which have an intrinsically $\delta$-like spectral appearance \cite{salkola_spectral_1997}, will be strongly influenced by $P(E)$-broadening.

In summary, we have shown that the interaction of tunneling electrons with the environmental impedance as well as the capacitative junction noise limit the effective energy resolution in spectroscopic measurements of the differential conductance. The $P(E)$-function that models the energy exchange with the electromagnetic environment combined with the capacitative noise is the energy resolution function of the tunnel junction. The effect of $P(E)$-broadening becomes dominant at or below 1\,K and has to be taken into account when optimizing the energy resolution. In this regime, the quantum nature of the tunneling process becomes evident, there is virtually no elastic tunneling, and the surrounding electromagnetic environment has to be taken into account.


\onecolumngrid
\newpage
\begin{center}
\textbf{\large Supplementary Information}
\strut
\vspace{1em}
\end{center}
\twocolumngrid
\section{Tip and Sample Preparation}

The experiments were carried out in an STM operating at a base temperature of 15\,mK \cite{assig_10_2013}. The sample that was used was an Al(100) single crystal and the tip was an Al wire of 99.9999\% purity. The sample was sputtered (Ar$^+$ ions at 500\,eV) and annealed in ultrahigh vacuum (low $10^{-10}$\,mbar range) in several cycles, while the tip was cut in air, transferred in vacuum and then sputtered (Ar$^+$ ions at 500\,eV) to remove the native oxide. With a superconducting transition temperature $T_C=1.1$\,K both tip and sample are superconducting at 15\,mK with a fully open gap. The quasiparticle density of states of aluminum in the superconducting state has a very BCS-like character with minimal intrinsic broadening \cite{giaever_tunneling_1962}, which is why aluminum is an excellent material for demonstrating the broadening effects in a tunnel junction.

\section{Modeling the Impedance $Z_T(\omega)$}

In our STM the surrounding impedance that contributes to the $P(E)$-function is the vacuum as well as the tip acting as a monopole antenna with a corresponding resonance spectrum that depends on the length of the tip \cite{jack_nanoscale_2015}. Approximating the resonance spectrum by an infinite transmission line impedance \cite{jack_nanoscale_2015,ingold_cooper-pair_1994}, we find an analytic expression for the impedance $Z(\omega)$:
\begin{equation}
Z(\omega) = R_{\text{env}} \frac{1+\frac{i}{\alpha}\tan\left(\frac{\pi}{2}\frac{\omega}{\omega_0}\right)}{1+i\alpha\tan\left(\frac{\pi}{2}\frac{\omega}{\omega_0}\right)}
\end{equation}
where $R_\text{env}$ is the effective dc resistance of the environmental impedance, $\alpha$ is an effective damping parameter, and $\omega_0$ is the frequency of the principal resonance. The parameter $R_\text{env}$ is set to the vacuum impedance of 376.73\,$\Omega$. The fit parameters for this impedance are $\alpha$ and $\omega_0$.

The total impedance $Z_T(\omega)$ takes into account the capacitance $C_J$ in the tunnel junction as well:
\begin{equation}
Z_T(\omega)=\frac{1}{i\omega C_J + Z^{-1}(\omega)}.
\label{eq:zt}
\end{equation}
Here, the parameter $C_J$ is also a fit parameter.

\section{Calculating the $P(E)$-function}

The $P(E)$-function is commonly defined through the equilibrium phase correlation function $J(t)$ through \cite{ingold_cooper-pair_1994}:
\begin{equation}
P(E)=\int\limits^{\infty}_{-\infty}\frac{dt}{2\pi\hbar}\exp[J(t)+iEt/\hbar]
\end{equation}
We regard the energy exchange with the environmental impedance and the capacitative noise from the tunnel junction as two independent processes, which allows us to separate the correlation function as:
\begin{equation}
J(t)=J_0(t)+J_N(t)
\end{equation}
where $J_0(t)$ is the phase correlation function from the environmental impedance and $J_N(t)$ is due to the capacitative junction noise. We can then calculate the corresponding probability functions separately, where $P_0(E)$ is the probability due to the interaction with the environmental impedance and $P_N(E)$ is due to the capacitative noise. Exploiting the convolution theorem, we can calculate the total $P(E)$-function through a convolution:
\begin{equation}
P(E)=\int\limits^{\infty}_{-\infty}dE'P_0(E-E')P_N(E')
\end{equation}
For the calculation of the $P_0(E)$-function for the environmental impedance, we follow the implementation given in Ref.\ \cite{ingold_finite-temperature_1991}. The $P_0(E)$-function is calculated through an indirect definition within an integral equation:
\begin{equation}
P_0(E)=I(E)+\int\limits^{\infty}_{-\infty}d\omega K(E,\omega)P_0(E-\hbar\omega)
\label{eq:PoEintegral}
\end{equation}
where $K(E,\omega)$ is the integral kernel. The inhomogeneity $I(E)$ is defined as:
\begin{equation}
I(E)=\frac{1}{\pi}\frac{D}{D^2+E^2}
\end{equation}
with
\begin{equation}
\quad D=\frac{\pi}{\beta}\frac{\Re Z_T(0)}{R_Q}
\end{equation}
where $\beta=(k_BT)^{-1}$, $T$ is the temperature, and $R_Q=h/(2e^2)$ is the resistance quantum. The integral kernel $K(E,\omega)$ is defined as:
\begin{equation}
   K(E,\omega)  =  \frac{\hbar E}{D^2+E^2}k(\omega) + \frac{\hbar D}{D^2+E^2}\kappa(\omega)
\end{equation}
with the functions $k(\omega)$ and $\kappa(\omega)$ being:
\begin{eqnarray}
     k(\omega) & = & \frac{1}{1-e^{-\beta\hbar\omega}}\frac{\Re Z_T(\omega)}{R_Q} - \frac{1}{\beta\hbar\omega}\frac{\Re Z_T(0)}{R_Q}\\
\kappa(\omega) & = & \frac{1}{1-e^{-\beta\hbar\omega}}\frac{\Im Z_T(\omega)}{R_Q} - \nonumber \\
& & \frac{2}{\beta\hbar}\sum\limits^\infty_{n=1}\frac{\nu_n}{\nu_n^2+\omega^2}\frac{Z_T(-i\nu_n)}{R_Q}\\
\nonumber
\end{eqnarray}

\begin{figure}
\centerline{ \includegraphics[width = 0.95\columnwidth]{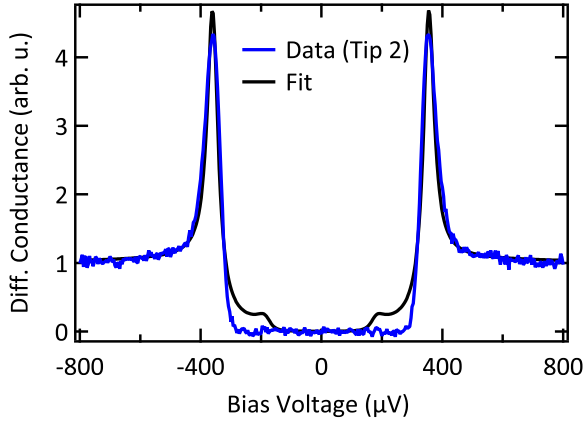}}
\caption{Differential conductance spectrum of tip 2 (blue) fitted by a Dynes function (black) disregarding the $P(E)$-broadening. The reduction of the singularity in the coherence peaks to finite values results in high $\Gamma$-parameters, which also fill the gap resulting in an unrealistic fit.} \label{fig:dynes}
\end{figure}

The Matsubara frequencies $\nu_n$ are defined as $\hbar\nu_n=2n\pi/\beta$. Using the inhomogeneity $I(E)$ as a starting value for the $P_0(E)$-function calculation, the integral equation Eq.\ \ref{eq:PoEintegral} can be solved self-consistently. Convergence is usually reached within a few iterations. Treating the integral as a convolution, the calculation can be done very efficiently numerically. Care should be taken to extend the integral range to sufficiently large energies, while at the same time having a high enough numerical point density. Other than the impedance $Z_T(\omega)$, the temperature $T$ is a fit parameter in this part of the $P(E)$-function calculation.

The $P_N(E)$-function for the thermal capacitative noise of the tunnel junction has proven a non-negligible part of the total $P(E)$-function. The low frequency capacitative noise $P_N(E)$ is modeled by a Gaussian \cite{ingold_cooper-pair_1994}:
\begin{equation}
P_N(E)=\frac{1}{\sqrt{4\pi E_C k_BT}}\exp\left[-\frac{E^2}{4E_Ck_BT}\right]
\end{equation}
where $E_C=Q^2/2C_J$ is the charging energy for Cooper pairs ($Q=2e$). The $P_N(E)$-function does not introduce any new fit parameters as the junction capacitance $C_J$ as well as the temperature $T$ are already defined in the $P_0(E)$-function.

\section{Fit parameters}

In the present work, we find $\hbar\omega_0=233\pm10\,\mu$eV, $\alpha=0.7\pm0.02$, and $C_J=3.5\pm0.2$\,fF for tip 1 as well as $\hbar\omega_0=120\pm15\,\mu$eV, $\alpha=0.75\pm0.05$, and $C_J=7.0\pm0.1$\,fF for tip 2. The fitted temperatures were $65\pm5$\,mK and $92\pm2$\,mK for tip 1 and 2, respectively. They agree well with the corresponding parameters in Ref.\ \cite{jack_nanoscale_2015}.

\section{Dynes fit}

In Fig.\ \ref{fig:dynes}, the differential conductance spectrum of tip 2 has been fitted with the Dynes equation \cite{dynes_direct_1978}:
\begin{equation}
n_{\text{t,s}}(E)=n_0\Re\left[\frac{E+i\Gamma}{\sqrt{(E+i\Gamma)^2-\Delta_{\text{t,s}}^2}}\right]
\end{equation}
where the phenomenological parameter $\Gamma$ introduces a general broadening of the BCS density of states. The fit neglects the $P(E)$-broadening, which means that the reduction of the singularities to finite values has to be absorbed in the $\Gamma$-parameter. Consequently, the broadening results in a rather large value $\Gamma=10.8\pm0.5\,\mu$eV. At the same time, there is a sizeable filling of the gap, which makes the Dynes model unsuitable for fitting these spectra. In this sense, we attribute the much better fit using the $P(E)$-function broadening also to the inherent asymmetry of the $P(E)$-function strengthening the validity of our approach.


\begin{thebibliography}{10}
\bibitem{li}
The lock-in modulation for the spectra was 10\,$\mu$eV at 724\,Hz. The modulation is small compared to the $P(E)$-broadening and can, therefore, be neglected in the context here.
\end{thebibliography}

\end{document}